\documentclass[11pt]{article}
\usepackage[a4paper,left=1in,right=1in,top=1in,bottom=1in]{geometry}
\usepackage{graphicx,amsmath,bm}
\usepackage{epsfig,subfigure}

\newcommand{\av}[1]{\langle #1 \rangle}

\title{Microscopic thermal machines using run-and-tumble particles}
\author{Aradhana Kumari\footnote{Email: aradhanakumari2546@gmail.com}~ and Sourabh Lahiri\footnote{Email: sourabhlahiri@gmail.com}}

\date{}

\begin{document}

\maketitle

\begin{center}
 \small Department of Physics, Birla Institute of Technology Mesra, Ranchi, Jharkhand 835215 
\end{center}

\begin{abstract}
Microscopic thermal machines that are of the dimensions of around few hundred nanometers have been the subject of intense study over the last two decades. Recently, it has been shown that the efficiency of such thermal engines can be enhanced by using active Ornstein-Uhlenbeck particles (AOUP). In this work, we numerically study the behaviour of tiny engines and refrigerators that use an active run-and-tumble particle (RTP) as the working system. We find that the results for the engine mode are in sharp contrast with those of engines using AOUP, thus showing that the nature of activity has a strong influence on the qualitative behaviours of thermal machines for nonequilibrium cycles. The efficiency of an engine using a run-and-tumble particle is found to be smaller in general than a passive microscopic engine. However, when the applied protocol is time-reversed, the resulting microscopic refrigerator can have a much higher coefficient of performance under these conditions. The effect of variation of different parameters of the coefficient of performance has been explored. A non-monotonic variation of coefficient of performance with active force has been found.
\end{abstract}


\section{Introduction}
Microscopic heat engines and refrigerators have been in the focus for the last two decades due to their possible applications across various fields. Micro-engines can find potential uses in the field of nanomachines and nanobots, which in turn would find applications in many areas, especially in medicine \cite{vyas2014}. 
An efficient microscopic engine needs to be developed in order to power such small machines. Several models have been proposed in the literature and have been realized experimentally, that would act as tiny heat engines \cite{sei08_epl,saha2019,lahiri2020,roldan2016,bechinger2012,lutz2016}. In a recent experiment \cite{sood2016},  using a trapped colloidal particle in a bacterial bath, the authors have demonstrated that the engine can attain much higher efficiencies than the one that uses normal thermal baths, even in the quasi-static limit.
A nice review on microscopic heat engines has been provided in \cite{roldan2017}. The extension to quantum dynamics has been done \cite{bender2000,quan2007}.

 It was shown in \cite{saha2019,lahiri2020} that an engine using an active Orntein-Uhlenbeck particle (AOUP) trapped in a harmonic potential as its working system turns out to be more efficient than its passive counterpart. The activity was introduced by means of an exponentially correlated noise that induced finite persistence in the motion of the particle. 
 Now, if such engines are driven in a time-reversed manner, then they can lead to microscopic refrigerators \cite{rana2016,pal2016}. Such refrigerators have also been modelled by using a sawtooth potential in presence of an external drive and have been studied in \cite{ai2006,lin2009,chen2010}. The basic principle of ordered motion in this case was studied in \cite{leo2010}.
 Refrigeration in presence of velocity-dependent feedback control has been studied in \cite{liang2000,kim2004,kim2007}. In \cite{liang2000}, the authors demonstrated how the accuracy of an AFM cantelever can be enhanced by reducing its thermal noise. Another potential application is in intramolecular cooling which, if achieved, can be beneficial in selecting certain reaction channels over others or for increasing the efficiency of catalysis \cite{arxiv2008}. Study of self-contained microscopic heat engines and refrigerators have been done for quantum systems \cite{linden2010,brunner2014}. 
 
 In this work, we use an active run-and-tumble (RT) particle as the working system of the engine and the refrigerator, unlike the active Ornstein-Uhlenbeck particle used in \cite{lahiri2020}. The RT particle is placed in a harmonic trap, whose stiffness parameter is varied cyclically with time, so as to mimic the steps of a macroscopic Stirling engine \cite{bechinger2012}.
RT particles provide an alternative model for activity, which is considered to be a more realistic model for describing motion of flagellar cells like the E. Coli in a stimulus-free environment \cite{berg1972}. They swim forward while their flagella spin counter-clockwise. However, once the direction of spin becomes clockwise, the flagella unbundle and the cell undergoes a rotation about its centre of mass. In \cite{condat2005,buceta2017,buceta2018}, the authors have provided  Langevin equations that capture the statistics of such motion. An effective Smoluchowski equation for RT motion in presence of taxis effects was developed in \cite{schnitzer1993}.
The RT motion in one-dimension has been modelled by a Langevin equation involving a telegraphic noise in \cite{dhar2019,chaki2019}. 
The dynamics has been contrasted with that of the active Brownian particles (ABPs) in \cite{cates2013}. For a recent review on the physics of such microswimmers, see \cite{elgeti2015}. Recently, it has been shown that RT particles can be prepared artificially by using synthetic Janus particles subjected to an electric field \cite{sano2017}.

We explore the behaviour of the RT particle in order to study its effects on the performance of thermal engines and refrigerators. An enhanced performance as compared to the passive engine/refrigerator should be of considerable practical significance. 
At first we show that the results of our simulations are in agreement with the analytical results of \cite{soto2020,condat2005}. Then we use simulations to study the system under the more general conditions where the stiffness parameter of the trapping potential changes periodically with time.
In addition to the usual RT motion, the particle is subjected to thermal noise due the presence of the two heat baths at different temperatures that are required for the working of the engine/refrigerator. We closely follow the approach of \cite{condat2005} in modelling our dynamics for an active particle in a thermal environment. The stochasticity of the system entails that the particle behaves in an engine mode or refrigerator mode only in a suitably chosen range of parameters \cite{saha2019,rana2016}.  We observe that although the engine mode is generally less efficient than that of the passive engine, the same system driven under a refrigerator protocol can yield enhanced performance. This has qualitative similarity with \cite{wijland2017_entropy}, where non-Gaussian noise was used to model activity. This implies that the nature of activity has a strong influence of the behaviour of the qualitative behaviours of the thermal machines operating out of equilibrium.


In section 2, we describe the detailed model for the active engine, both in one and two dimensions. In section  3, we provide comparison of our numerical simulations of an RT particle with known analytical results, and then study numerical results for the efficiency of the active engines and their comparison with the corresponding passive ones. Section 4 describes the model for the refrigerator in one and two dimensions. Section 5 discusses the results for the active particle subjected to the refrigerator protocol in one and two dimensions. Finally, the conclusions are provided in section 6.

\section{The Engine Model} \label{sec:eng_model}

\begin{figure}[!ht]
\centerline{\psfig{file=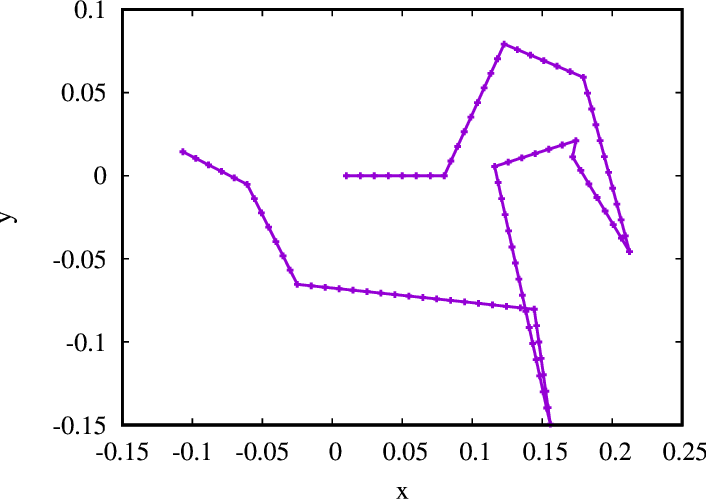,width=3.0in}}
\vspace*{8pt}
\caption{A typical trajectory of a 2d RT particle. The symbols represent individual time steps.}
\label{fig:traj2d}

\end{figure}

An RT particle proceeds through a series of ``runs'' followed by ``tumbles'', whose motion was first described in \cite{berg1972}. During the runs, the particle does not change direction and follows the path dictated by deterministic dynamics. However, at the end of a single run, the particle suddenly changes its direction randomly, and then sets into motion for another run along this newly chosen direction. The particle is thus said to tumble and choose a new direction in-between two run-times. Since the tumble times are typically an order of magnitude smaller than the typical run-times \cite{buceta2017}, we will approximate it to zero in our analysis. A typical trajectory of such a particle in absence of thermal noise is shown in figure \ref{fig:traj2d}, in the space of normalized position coordinates (see the discussion below Eq. \eqref{theory}). The symbols appear after every time step of integration. The persistence in motion over each runtime can be clearly observed.

The run times and tumble times have been shown to roughly follow  exponential distributions \cite{berg1972,berg1974}.
We therefore sample our run-times from the distribution 
\begin{align}
 P(\tau_r) &= \frac{1}{\tau_p} e^{-t/\tau_p},
 \label{expdist}
\end{align}
where $\tau_p\equiv \langle\tau_r\rangle$ is the average run-time, which we refer to as the persistence time. 
An exact mathematical description is lacking, although the asymptotic behaviour (in the limit of large time) can be shown to reach the diffusive regime \cite{soto2020}. Tumbling characteristics can broadly be of two types: 
\begin{enumerate}
 \item The particle undergoes complete reorientation (it can  choose all possible deflection angles with equal probability).
 
 \item The particle undergoes partial reorientation, where the probabilities of larger deflections are smaller. In particular, we choose the cosine distribution (see Eq. \eqref{cosine_distribution} below).
\end{enumerate}
Our dynamics consists of the RT motion of the particle, as well as the thermal motion generated due the finite temperature of the medium in which the particle is present. We now consider the one dimensional model in detail, and compare the results with the two dimensional one.

\subsection{1d model for RT engine}
\label{sec:1d model}

The particle experiences a potential $V(x)=k(t)x^2/2$ created by an optical trap. 
The harmonic trap undergoes four steps corresponding to a Stirling cycle \cite{bechinger2012,lahiri2020}, given below. Figure \ref{fig:stiffness}(a) shows the variation in the shape of the potential during the cycle, while functional form of $k(t)$ is shown in figure \ref{fig:stiffness}(b).
\begin{enumerate}
 \item {\bf Step 1 (isothermal expansion)}: the temperature is fixed at the value $T_h$ (corresponding to temperature of the hot thermal bath), and $k(t)$ varies linearly with time from the initial stiffness value $k_0$ to the final value $k_0/2$:
 \begin{align}
  k(t) &= k_{\mathrm{exp}}(t) =  k_0\left(1-\frac{t}{\tau}\right), \hspace{0.5cm} 0<t\le \tau/2
 \end{align}
 The randomness in the dynamics is caused by the thermal noise of strength $D_h = 2\gamma k_B T_h$, where $\gamma$ is the friction coefficient of the medium, and $k_B$ is the Boltzmann constant.
 
 \item {\bf Step 2 (isochoric temperature decrease):} The value of stiffness parameter is held constant, and the temperature is suddenly changed (in comparison to the sampling frequency, as per the set-up used in \cite{bechinger2012}) to the lower value $T_c$.
 
 \item {\bf Step 3 (isothermal compression):} The temperature is held constant at $T_c$, while the stiffness parameter changes from $k_0/2$ to $k_0$ as per the following linear function:
 \begin{align}
  k(t) &= k_{\mathrm{com}}(t) = k_0\left(\frac{t}{\tau}\right), \hspace{1cm} \tau/2<t\le \tau.
 \end{align}
 In this step, in addition to activity, a thermal noise of strength $D_c=2\gamma k_B T_c$ is present due to the finite temperature of the medium.
 
 \item {\bf Step 4 (isochoric temperature increase):} In the final step, the temperature is suddenly increased to $T_h$, keeping the stiffness parameter fixed, so as to allow the engine to complete a cycle.\\

\end{enumerate}

\begin{figure}[!ht]
\centering     
\subfigure{\psfig{file=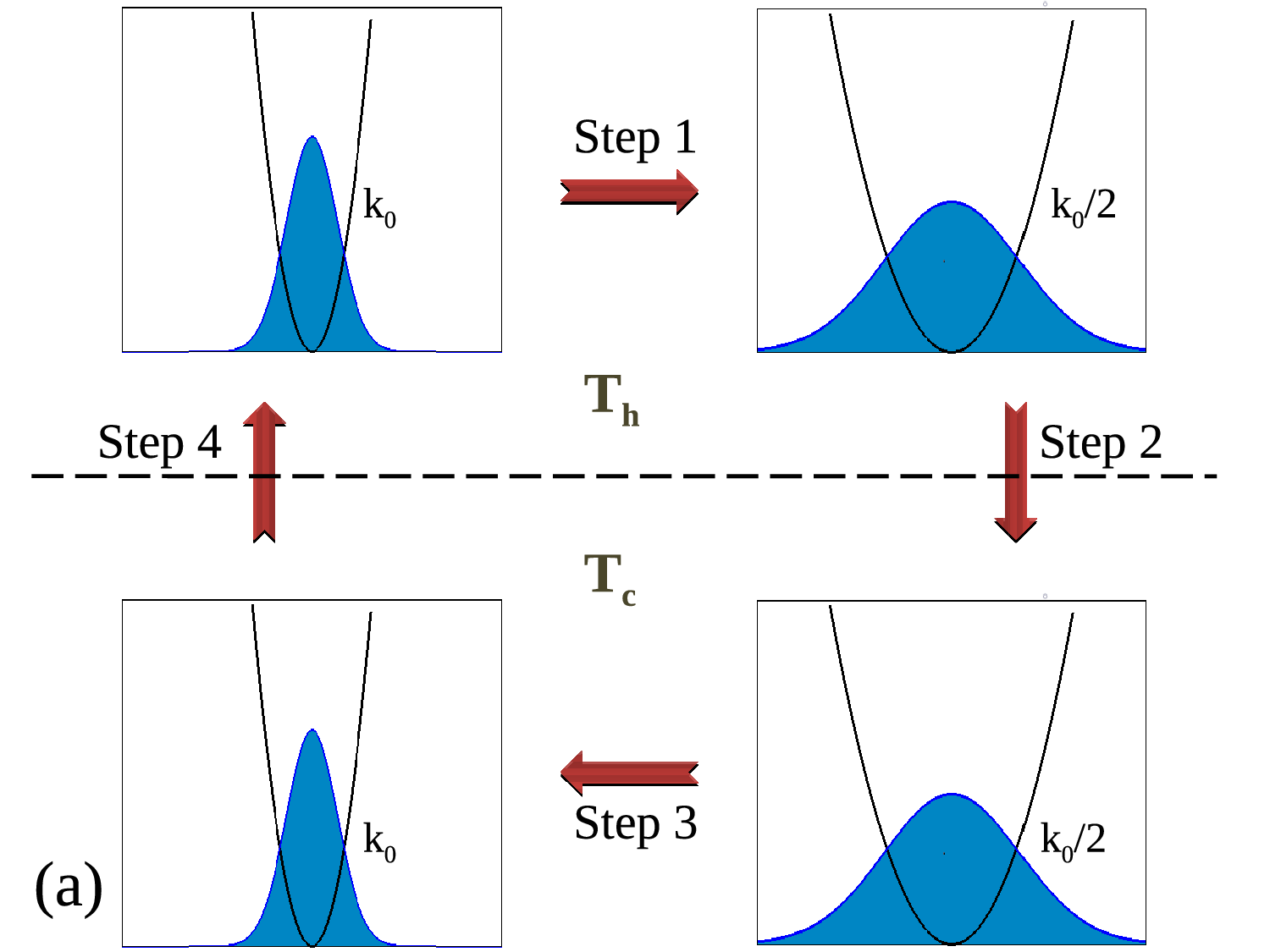,width=2.4in}}
\hfill
\subfigure{\psfig{file=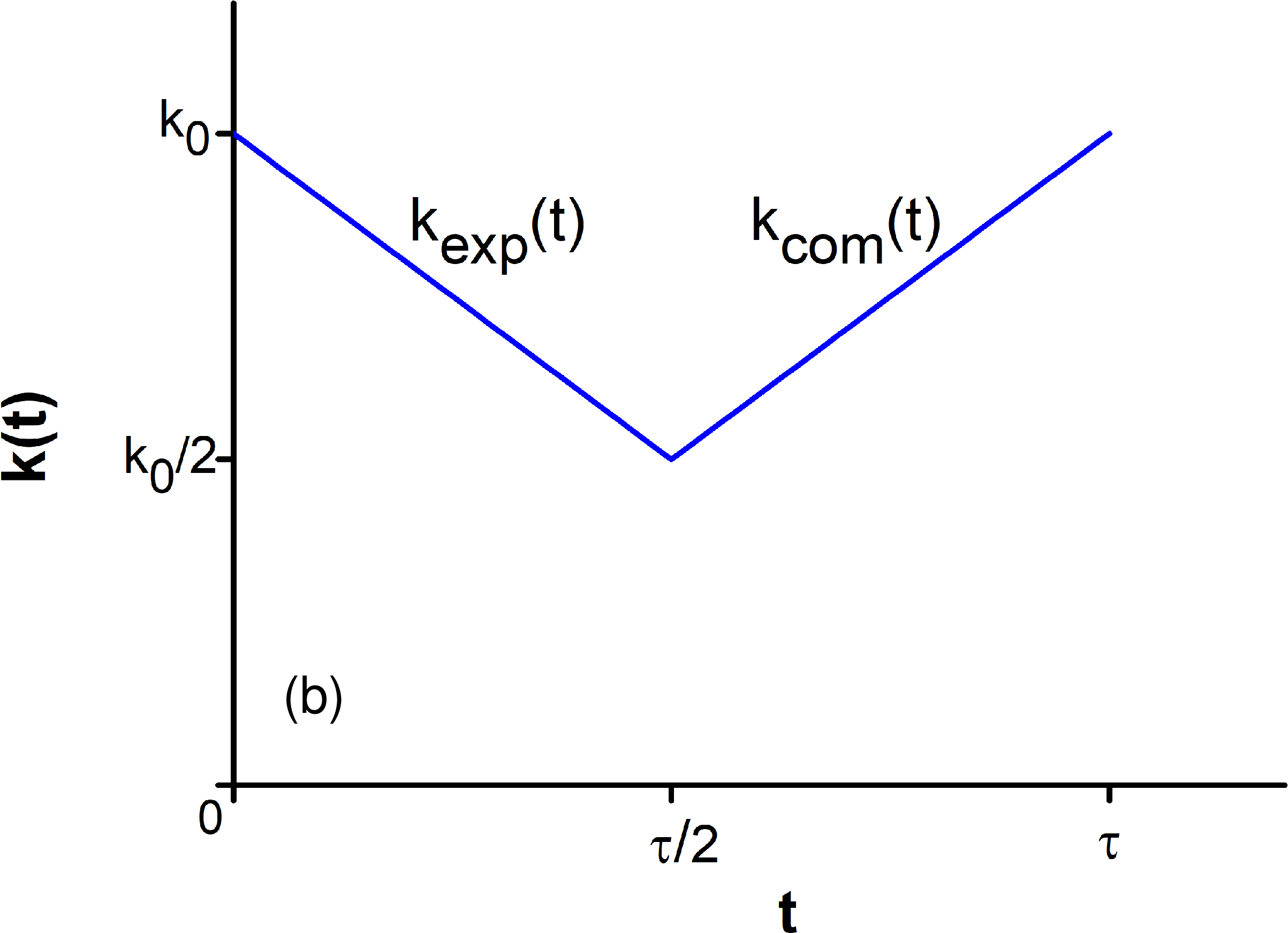,width=2.4in}}
\vspace*{8pt}
\caption{(a) Stirling cycle under engine protocol. (b)Variation of stiffness parameter with time in the engine protocol.}
\label{fig:stiffness}

\end{figure}

In the case of total reorientation, the particle moves either along the $+x$ or $-x$ direction for a time duration $\tau_r$, and then suddenly changes direction ($+x\to -x$ or vice versa) with probability 1/2. The persistence time is chosen to be typically much smaller than the cycle time, so that each isothermal step consists of several run-and-tumble events. When the particle undergoes partial reorientation, it switches direction more reluctantly, with a probability $p<1/2$. With the switch in direction, the active force also changes sign.

During the expansion step, in order to make comparisons with our results in \cite{lahiri2020}, we keep the particle inactive. This can be achieved, for instance, by photoactivation \cite{buttinoni2012,uspal2019}.
The equation of motion is given by (see \cite{condat2005})
\begin{align}
 m\dot v &= -\gamma v - k_{\rm exp}(t)x + (\sqrt{D_h})\xi(t), \hspace{1cm} 0\le t<\tau/2
 \label{1d_RT_exp}
\end{align}
where $\xi(t)$ is a normally distributed white noise: $\langle\xi(t)\rangle=0$, $\langle\xi(t)\xi(t')\rangle=\delta(t-t').$ In a few cases, however, we will compare the results of this model with the one in which the activity is present throughout the circle.

During the compression step, the equation of motion during the $i^{th}$ run-time is given by
\begin{align}
 m\dot v_i &= -\gamma v_i - k_{\rm com}(t)x + F + (\sqrt{D_c})\xi(t),\nonumber\\
 &\hspace{1.5cm} \tau/2 \le t_i \le t < t_i+\tau_{r,i}\le\tau.
 \label{1d_RT_com}
\end{align}
$\tau_{r,i}$ represents the $i^{th}$ run-time, while the active force $F$ propels the particle in the chosen direction during a run \cite{condat2005}. Here, $t_i$ is the initial time for the $i^{th}$ run, which ends at time $t_i+\tau_{r,i}$. If activity is present in the expansion half as well, then a similar equation of motion will take the place of Eq. \eqref{1d_RT_exp}.

\subsection{2d model}
\label{sec:2d_model}

In this case, the particle is trapped in a 2d harmonic potential, $V(x,y)=k(t)(x^2+y^2)/2$. We can describe the position vector of the particle by ${\bf r}=x\hat x+y\hat y$ during a run-time ($\hat x$ and $\hat y$ are the unit vectors in the $x$ and $y$ directions, respectively), which changes to ${\bf r}'=x'\hat x+y'\hat y$, and the angle $\theta$ between ${\bf r}$ and ${\bf r}'$ is the deflection attained due to the sudden tumble at the end of a run-time. 

For the simple case of total reorientation, we use a uniform random number $\eta$ at each tumble event in the range $\eta\in[-\pi,\pi]$, and allow the value of the angle (with the $x-$axis, for convenience) $\theta$ to change to $\theta+\eta$. 
For the case of partial reorientation, we choose the distribution of the deflection to be of the form 
\begin{align}
P(\eta)=\frac{1}{4}\cos\left(\frac{\eta}{2}\right), 
\label{cosine_distribution}
\end{align}
where the range is $\eta\in[-\pi,\pi]$. The method of obtaining the distribution has been given in Appendix A.
The stiffness parameter $k(t)$ undergoes the same set of steps as mentioned in section \ref{sec:1d model}.

Let the total velocity be given by ${\bf v}=v_x\hat x + v_y\hat y$. Then the equations of motion are given by 
\begin{align}
 m\dot v_{x} &= -\gamma v_{x} - k_{\rm exp}(t)x + (\sqrt{D_h})\xi(t), \hspace{0.5cm} 0<t\le\tau/2;\nonumber\\
 m\dot v_{x,i} &= -\gamma v_{x,i} - k_{\rm com}(t)x + F_x + (\sqrt{D_c})\xi(t), \nonumber\\
 &\hspace{3.5cm} \tau/2<t_i< t \le t_i+\tau_{r,i}\le\tau,
 \label{2d_RTmodel}
\end{align}
where $\tau_{r,i}$ is the $i^{th}$ run-time. ${\bf F}$ is a self-propelling force that is directed along the axis of the particle, having components $F_x=F\cos\theta$ and $F_y=F\sin\theta$. Similar equations can be written for motion in the $y$-direction. 
\section{Results and Discussions for stochastic engine using RT particle}
\label{sec:active_engine}

\subsection{Comparison with theory}
 
We first test the outcomes of our simulation against theory. It was shown in \cite{soto2020} that the mean squared displacement of a two-dimensional RT particle (in the absence of any thermal bath or any potential) leads to the relations
\begin{align}
 \langle r^2\rangle_{\mathrm{tot}} &= 2v_0^2\tau_p^2\left(\frac{t}{\tau_p} + e^{-t/\tau_p}-1\right);\nonumber\\
 \langle r^2\rangle_{\mathrm{par}} &= \frac{2v_0^2\tau_p^2}{1-\sigma_1}\left[\frac{t}{\tau_p} + \frac{e^{-(1-\sigma_1 )t/\tau_p}-1}{1-\sigma_1}\right].
 \label{theory}
\end{align}
Here,  $\sigma_1 = \langle\cos(\eta)\rangle$ is the Fourier transform of the tumbling kernel (see \cite{soto2020}). The parameter $v_0$ is the constant magnitude of the total velocity. On the other hand, in presence of thermal noise, the change in steady-state variance of an RT particle (without any potential) can be obtained using the approach of \cite{condat2005}. If the friction coefficient of the thermal bath is $\gamma$ and its thermal noise strength is $D$, while the self-propelling force $F$ is set to zero for analytical convenience, then we arrive at the expression (see Appendix B for details)
\begin{align}
 \langle r^2(t)\rangle &= \left[\frac{2D\tau_p}{\gamma(m+\gamma \tau_p)}\right]t + \left[\frac{2Dm\tau_p^2}{\gamma(m+\gamma\tau_p)^2}\right]\nonumber\\
 &\times\exp\left\{-\frac{(m+\gamma\tau_p)t}{m\tau_p}\right\} - \frac{2Dm\tau_p^2}{\gamma(m+\gamma\tau_p)^2}.
 \label{var_condat}
\end{align}

\begin{figure}[!ht]

\centering     
\subfigure{\psfig{file=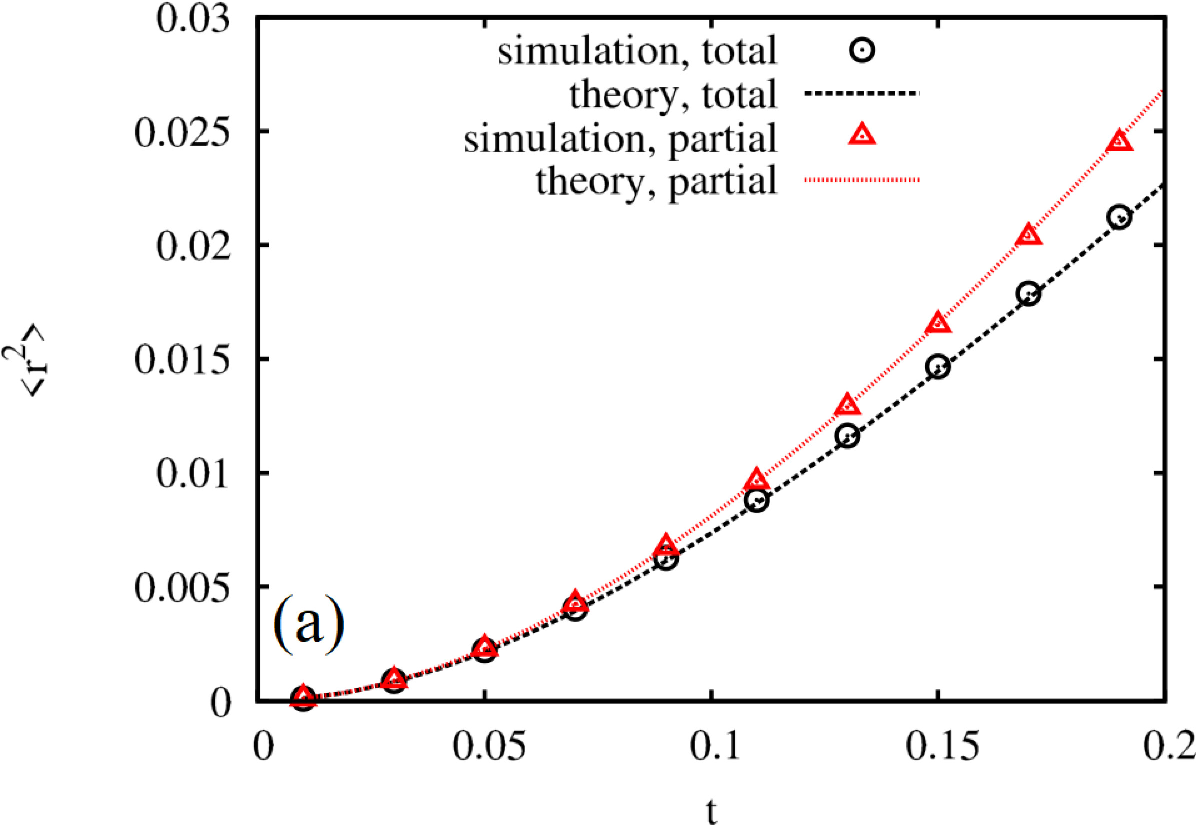,width=2.4in}}
\subfigure{\psfig{file=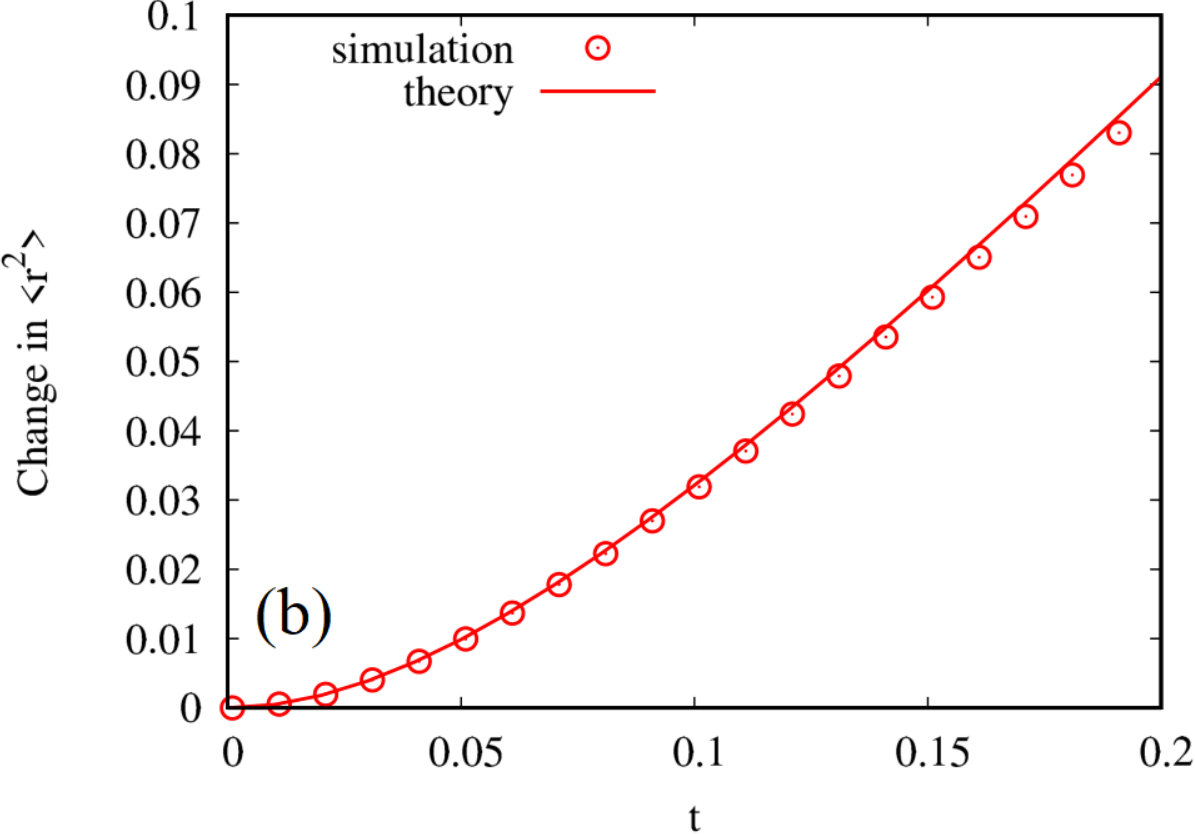,width=2.4in}}
\vspace*{8pt}
\caption{(a) Plots of the analytical solution in comparison to the numerical results for total reorientation (black solid line and black open circles, respectively) and for partial reorientation (red solid line and red open triangles respectively) of a 2d RT particle. The parameters are: $m=0.2, ~\tau_p=0.1$.(b) Plots for the change in variance of the particle as a function of time in the steady state when thermal noise is present. The parameters are $m=0.2,~\tau_p=0.1,~D=1$.}
\label{fig:analytics}
\end{figure}

We have shown these results in figure \ref{fig:analytics}, where the simulation results have been obtained by using the Heun's method of integration (see \cite{man2000}) and by averaging over $\sim 10^5$ realizations. In the figures, all times are normalized by the cycle time $\tau$. All positions are normalized by the factor $\sqrt{D_c\tau}$ ($D_h$ and $D_c$ are the diffusion constants for the hot and the cold reservoirs, respectively). This leads to a characteristic mass scale of $\gamma\tau$ and characteristic force scale of $\gamma\sqrt{D_c/\tau}$. In figure \ref{fig:analytics}(a), equations \eqref{theory} have been compared with the results of simulations for total and partial reorientations. The symbols represent results of simulations, while the solid lines are the analytical results. 

Figure \ref{fig:analytics}(b) compares Eq. \eqref{var_condat} with simulation outcomes (single bath with a noise strength of $D$). In all the cases, the results of simulation are in good agreement with theoretical results. Unfortunately, these analytical treatments become intractable in presence of external time-dependent potentials. The nature of velocity correlations have been discussed in Appendix C.

\subsection{Efficiency of the 1d active RT engine}
	
	We now proceed with the calculation of efficiency of an engine using a 1d active particle by simulating the dynamics of the engine (see Eqs. \eqref{1d_RT_exp} and \eqref{1d_RT_com}). 	
	 The work done on the particle is calculated by using the definition as prescribed in Stochastic Thermodynamics \cite{sek98,sekimoto}:
\begin{align}
 W(t) &= \frac{1}{2}\int_0^t \dot k(t')x^2(t')dt'. 
\end{align}
The average work extracted can be calculated from the variance, defined as $\sigma(t)=\langle x^2(t)\rangle$, by means of the relation
\begin{align}
 \langle W(t)\rangle &= \frac{1}{2}\int_0^t \dot k(t')\sigma(t')dt'.
\end{align}
The average heat absorbed in the expansion step is given by the first law:
\begin{align}
 \langle Q\rangle_h &= \langle W\rangle_h - \langle\Delta E\rangle_h,
\end{align}
where the subscript $h$ implies that only the expansion step (when system is in contact with hot reservoir) is considered. The sign convention is such that the work done \emph{on} the system and the heat dissipated \emph{by} the system are positive, which is why  the extracted work and absorbed heat must be given by $-\av W$ and $-\av{Q}_h$, respectively. The average change in internal energy  during the expansion step is given by
\begin{align}
 \langle\Delta E\rangle_h &= \frac{1}{2}k(\tau/2)\sigma(\tau/2) - \frac{1}{2}k(0)\sigma(0).
 \label{DeltaEh}
\end{align}
The efficiency is obtained using the standard relation
\begin{align}
 \eta &= \frac{\av W}{\av{Q}_h}= \frac{\av W}{\av{W}_h-\av{\Delta E}_h}.
\end{align}

\begin{figure}[!ht]
\centering     
\subfigure{\psfig{file=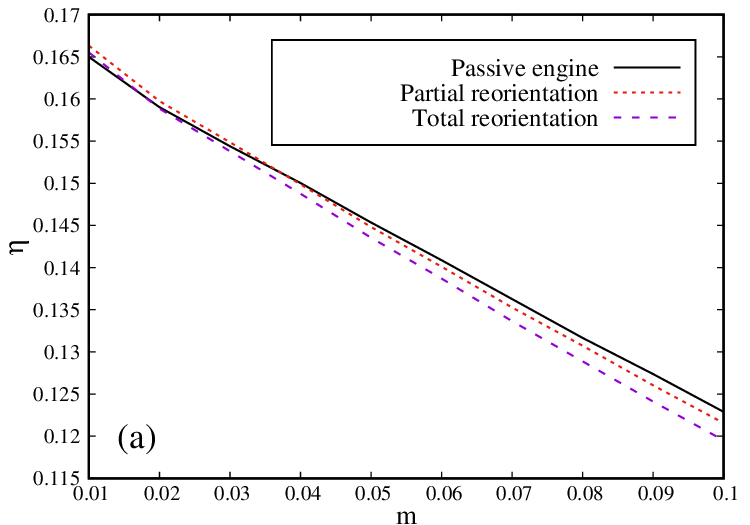,width=2.4in}}
\subfigure{\psfig{file=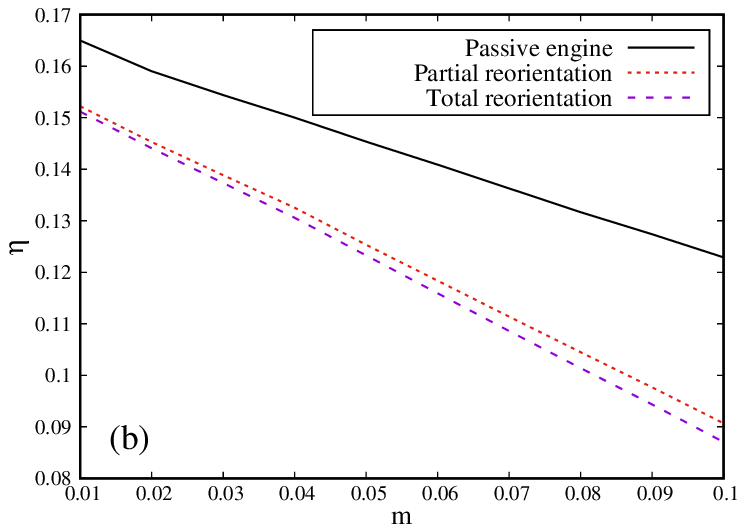,width=2.4in}}
\vspace*{8pt}
\caption{(a) Plot of efficiency with mass in 1d for the passive engine (black solid line), RT engine, with activity only in the compression cycle, with total reorientation (magenta dashed line) and with partial reorientation (red dotted line). The parameters are $D_{c}=1$, $D_{h}=2$, $k_0=50$, $\tau_p=0.2$, $F=0.1$.  (b)Similar plot when the activity is present in the full engine cycle.}
\label{fig:RT1_eff_m}

\end{figure}
We now compare the efficiencies of the active and passive engines, when the engine has settled into a time-periodic steady state, i.e. when the variance becomes periodic in time. We show this in figure \ref{fig:RT1_eff_m}(a), where we  observe the general trend that the efficiency of the passive engine is slightly higher than the active engine with partial reorientation, the latter being in turn slightly higher than the active engine with total reorientation. However, the differences are quite small in our case. 
In figure \ref{fig:RT1_eff_m}(b), we have plotted similar graphs for the case where activity is present throughout the full cycle, and the difference between the passive and active engines is more pronounced in this case. Consequently, we conclude that unlike in the case of AOUP (see \cite{lahiri2020}, and Appendix D), the passive particle turns out to produce a more efficient engine than an RTP. This is likely due to the degradation of persistence owing to the tumble events for smaller values of $F$, and the decrease in compressibility in the compression step for higher values of $F$. The general trend can be found to be a decreasing efficiency with increase in the mass of the particle. This is expected, because more work is to be pumped into the system during the compression cycle when the particle is more massive, thereby reducing the efficiency. We have separately checked the variation of efficiency with $\tau_p$ and with $F$, but in all cases the passive engine outperforms its active counterpart.

\subsection{Efficiency of a 2d RT engine}

We obtain similar trends even when the engine is two-dimensional and follows the equations of motion as outlined in section \ref{sec:2d_model}, namely the active engines are not able to outperform the passive ones.
Thus, given that the active engine does not provide any major advantage over its passive counterpart, we next focus our attention on the refrigerator protocol.

\section{The Refrigerator Model} \label{sec:ref_model}

\begin{figure}[!ht]
\centering     
\subfigure{\psfig{file=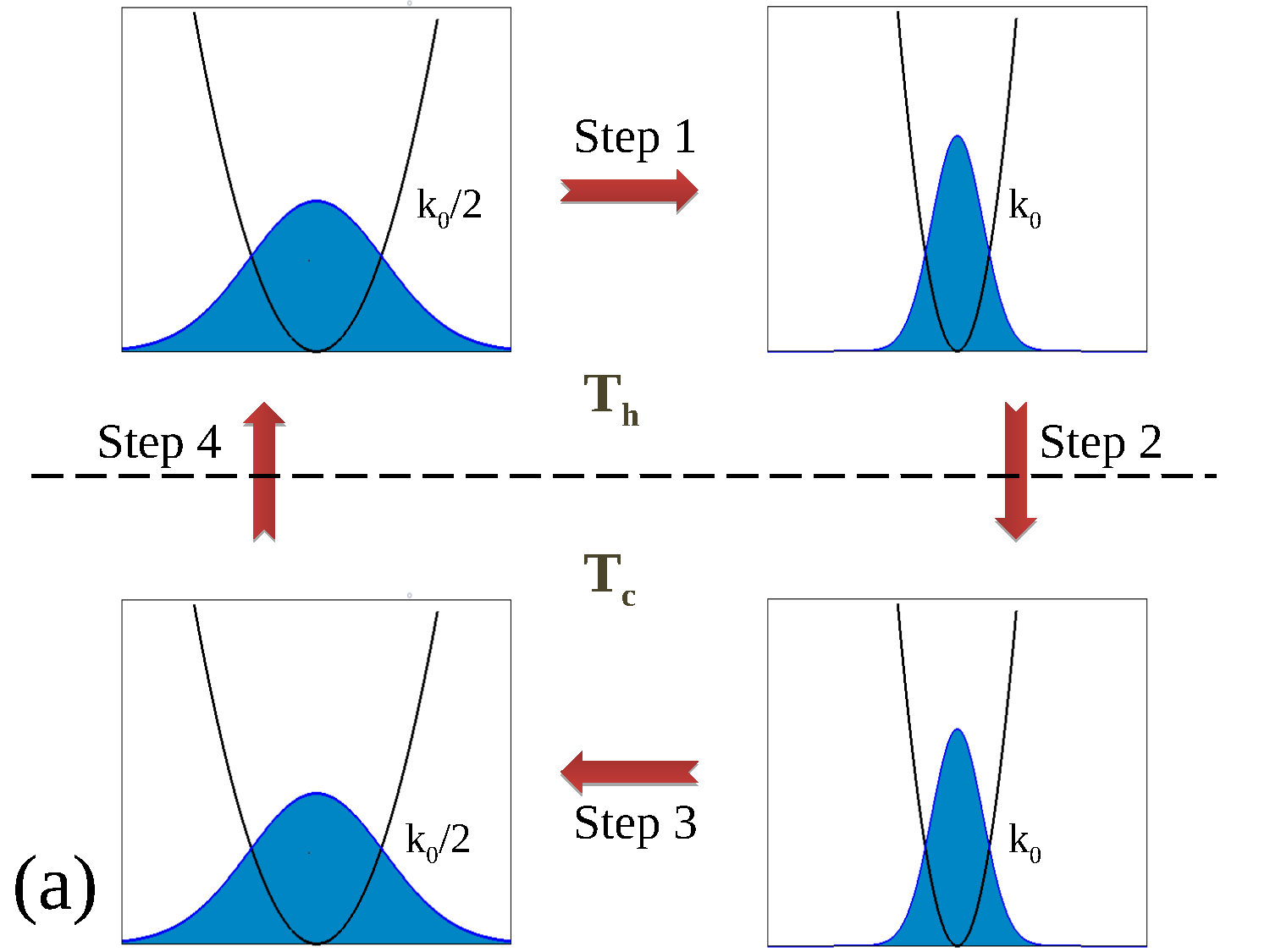,width=2.4in}}
\hfill
\subfigure{\psfig{file=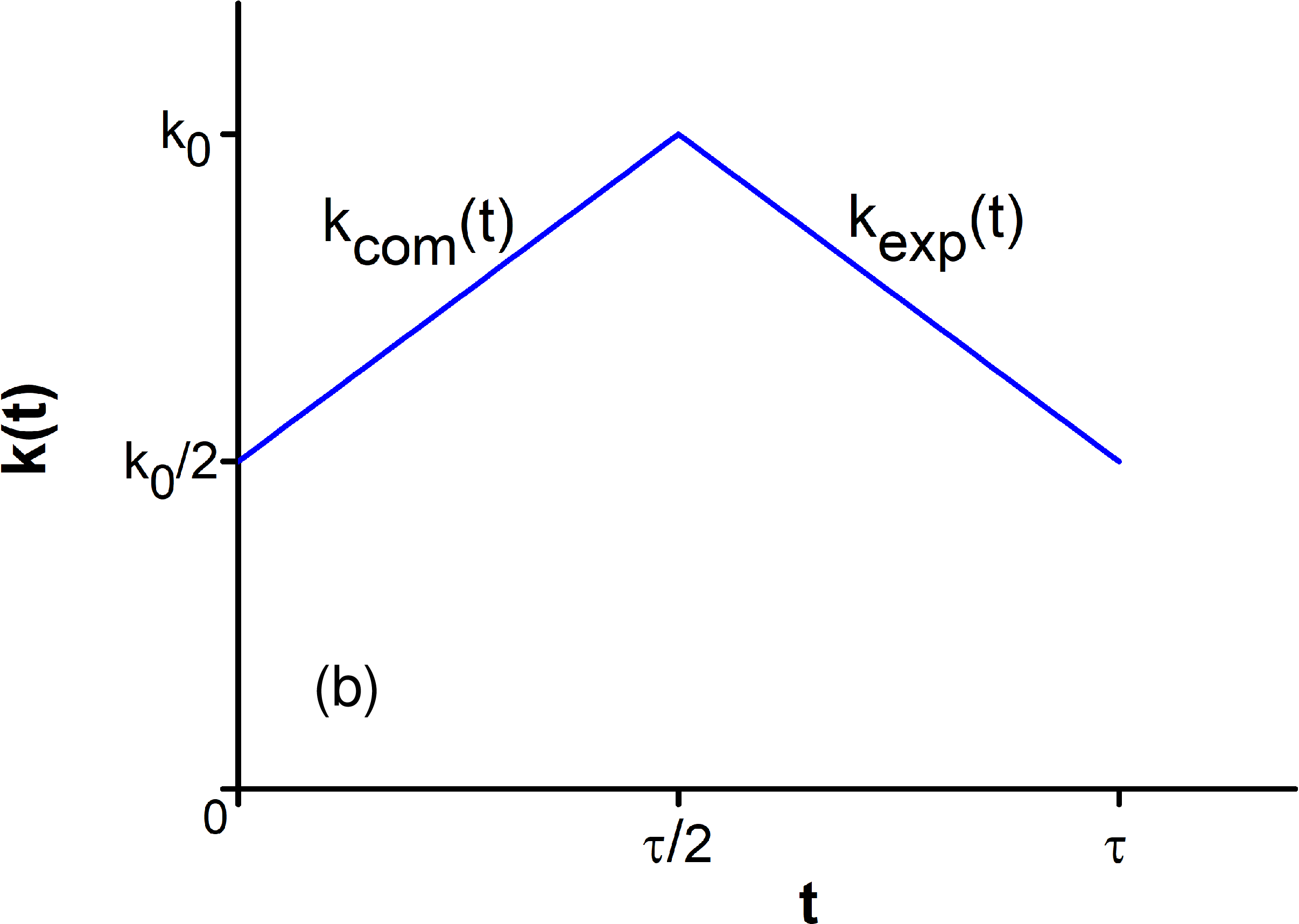,width=2.4in}}
\vspace*{8pt}
\caption{(a) Stirling cycle under refrigerator protocol. (b)Variation of stiffness parameter in the  refrigerator protocol.}
\label{fig:stiffness_ref}

\end{figure}

In this section, we study the behaviour of the RT particle undergoing partial reorientations in one and two dimensions, by reversing the engine protocol, as shown in figure \ref{fig:stiffness_ref}(b). 

The time-dependences are given by
\begin{align}
 k_{\mathrm{com}}(t) &= k_0\left(\frac{1}{2}+\frac{t}{\tau}\right), \hspace{1cm} 0< t\le \tau/2; \nonumber\\
 k_{\mathrm{exp}}(t) &= k_0\left(\frac{3}{2}-\frac{t}{\tau}\right) \hspace{1cm} \tau/2< t\le \tau.
\end{align}
The sequence of steps are given below. The equations of motion can be written down in the same way as in the engine protocol (see sec. \ref{sec:eng_model}).
\begin{enumerate}
 \item {\bf Isothermal compression} at temperature $T_h$, changing stiffness constant from $k_0/2$ to $k_0$. In this step, activity is switched on.
 \item {\bf Isochoric decrease of temperature} from $T_h$ to $T_c$. The stiffness parameter remains fixed at $k_0$.
 \item {\bf Isothermal expansion} at temperature $T_c$,  changing the stiffness constant from $k_0$ to $k_0/2$. The particle is inactive in this step.
 \item {\bf Isochoric increase of temperature} from $T_c$ to $T_h$. The stiffness parameter remains fixed at $k_0/2$.
\end{enumerate}

The \emph{coefficient of performance} (COP) of a refrigerator is defined as the ratio of the heat removed from the cold reservoir to the work done in the full cycle:
\begin{align}
 \mathrm{COP} &= -\frac{\av{Q}_c}{\av{W}} = -\frac{\av{W}_c-\av{\Delta E}_c}{\av{W}}.
\end{align}
The subscript $c$ refers to the fact that the quantities have been computed when the particle is in contact with the cold reservoir.

For a reversible engine, we find that
\begin{align}
 \mathrm{COP} &= \frac{T_c}{T_h-T_c} = \frac{T_c}{\eta_{\mathrm{rev}} T_h},
 \label{COP_rev}
\end{align}
where $\eta_{\mathrm{rev}} \equiv (1-T_c/T_h)$ is the Carnot efficiency in the engine mode. Thus, in the nonequilibrium regime, it would be reasonable to expect that a more efficient engine when run in reverse would give a refrigerator with smaller COP, and vice versa. Keeping this in mind, we study the thermodynamics of the active refrigerator and check whether it can have a higher value of COP than a passive refrigerator in a suitable range of parameters.

\section{Results and Discussions for Stochastic Refrigerator using RT particle} \label{sec:ref_discussions}
\subsection{COP of a 1d RT refrigerator}

\begin{figure}[!ht]
\centering     
\subfigure{\psfig{file=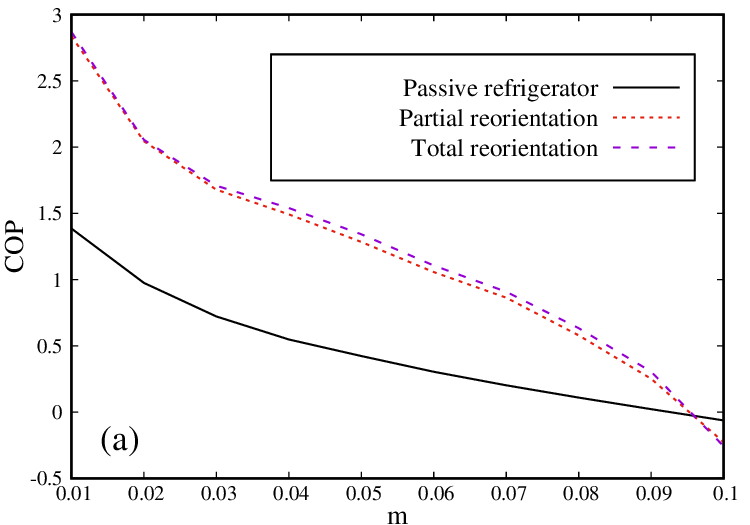,width=2.4in}}
\subfigure{\psfig{file=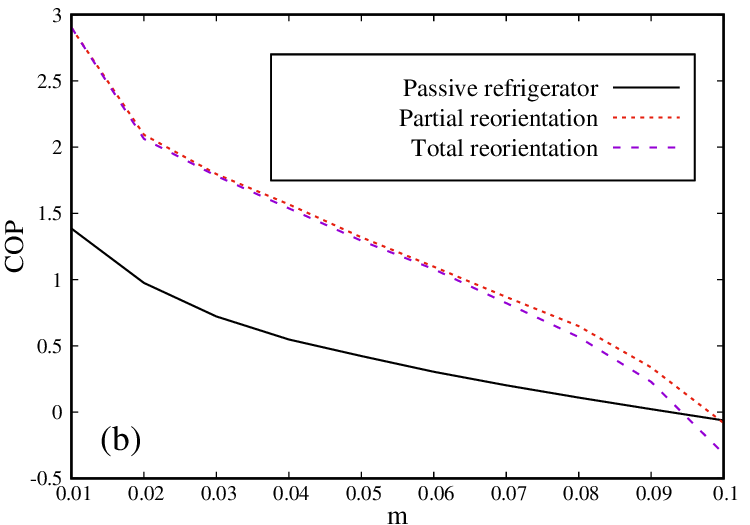,width=2.4in}}
\vspace*{8pt}
\caption{(a) Plot of the coefficient of performance of the 1d refrigerator with total (magneta dashed line) and partial (red dotted line) reorientation as a function of the mass of the particle. The black solid line is the COP for a passive refrigerator. The activity is present only in the compression step. The parameters are: $F=0.1$ (for active particle), $D_h=1.25,~D_c=1,~\tau_p=0.2$. (b)Similar plot when the activity is present in the full refrigerator cycle.}
\label{fig:RT1_cop_m}
\end{figure}

In figure \ref{fig:RT1_cop_m}(a), we plot the variation of the COP with the mass for 1d active (total as well as partial reorientation) and passive particles. For a suitable range of parameters (see figure caption), the active refrigerator is observed to have a higher COP than the passive one. We will focus only on the range where the system acts as a refrigerator ($\av{Q}_c<0,~\av{W}>0$), which happens when the mass of the particle is less than approximately 0.09 in the diagram, where the other parameters are as mentioned in the caption. For sake of comparison, we also show the behaviour of COP with mass in the case of a fully active cycle (i.e. the RT motion is present in both halves of the cycle), which show a similar trend as in (a). Since the partial reorientation is observed to produce little difference from total reorientation, we will henceforth deal only with a particle that undergoes a partial reorientation.

\begin{figure}[!ht]
\centering     
\subfigure{\psfig{file=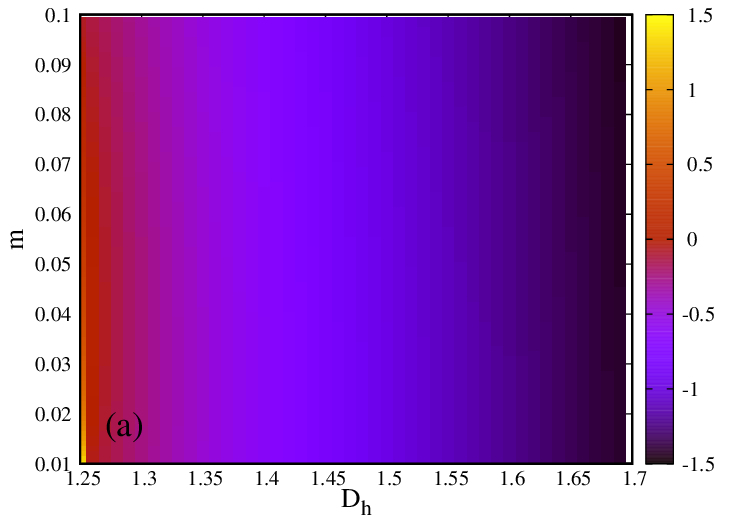,width=2.4in}}
\subfigure{\psfig{file=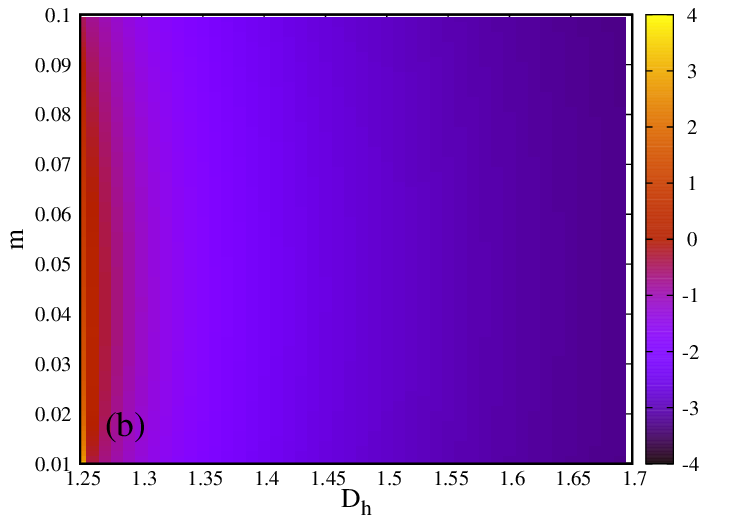,width=2.4in}}
\vspace*{8pt}
\caption{Phase plot of COP as a function of $D_h$ and $m$ for (a) 1d passive refrigerator(b)1d active refrigerator. Parameters are $D_c=1,~k_0=50,\tau_p=0.2$.}
\label{fig:ref_phaseplot}

\end{figure}

In figure \ref{fig:ref_phaseplot}, we show the phase plot indicating the variation of the COP of the passive  and active refrigerators (figures \ref{fig:ref_phaseplot}(a) and (b), respectively) as a function of the thermal noise $D_h$ of the hot bath and of the mass $m$ of the particle. One can observe the reduced value of COP when the thermal drive is smaller, in conformity with our expectations as discussed above (if engine is inefficient, COP tends to be higher). The highest values of COP is obtained when both $m$ and $D_h$ are small. Although we find similar qualitative trends for both passive and RT particles, the COP is observed to be larger in the latter case.
Overall, we find that although the engine using RT particle as its working substance finds it hard to beat the performance of a passive engine, the situation is quite different when one deals with the corresponding refrigerators by time-reversing the protocols. The RT refrigerator in a suitably chosen range of parameters can yield much better results than a passive refrigerator.

\begin{figure}[!ht]
\centering     
\subfigure{\psfig{file=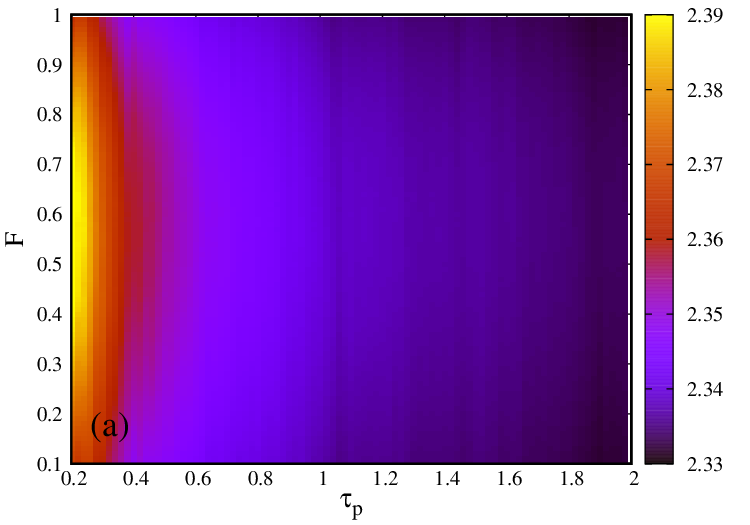,width=2.4in}}
\subfigure{\psfig{file=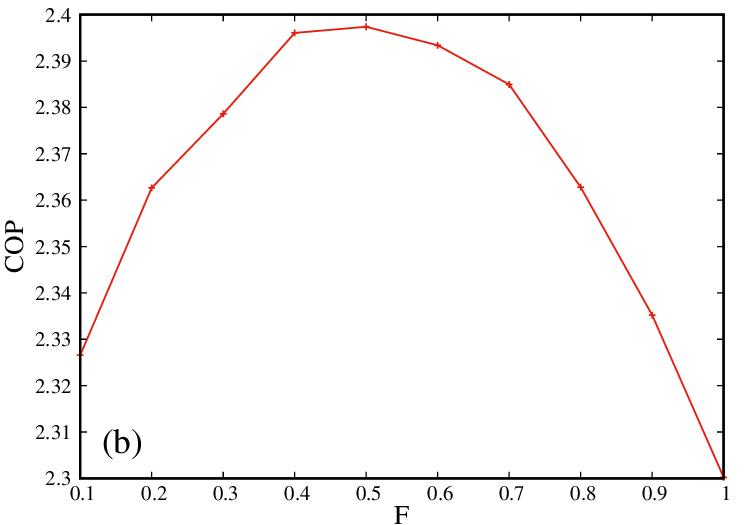,width=2.4in}}
\vspace*{8pt}
\caption{(a) Phaseplot of 1d active refrigerator, where the COP has been plotted as a function of the active force $F$ and the persistence time $\tau_p$. The parameters used are : $k_0=50,~D_c=1,~D_h=1.25,~m=0.01$.  (b) Variation of COP with $F$ for mass=0.01.}
\label{fig:comp_F_taup}

\end{figure}

Figure \ref{fig:comp_F_taup}(a) shows the variation in COP with the mean run-time $\tau_p$ and the active force $F$. We find that the COP generally decreases with increase in run-time. However, as a function of $F$, the behaviour turns out to be non-trivial at smaller values of $\tau_p$, where a non-monotonic variation of the COP can be observed. This region has been explored in figure \ref{fig:comp_F_taup}(b), where the maximum of COP is found roughly at around $F=0.5$, where the other parameters are as mentioned in the figure caption.

Thus, it is possible to tune the external parameters to make the refrigerator run at a high COP. The general rule would be to allow the system to function at relatively smaller values of mass and temperature difference, and control the value of the active force $F$, which turns out to be more important in determining the efficacy of performance of the refrigerator, in contrast to the persistence time $\tau_p$ that seems to produce a much smaller influence.

\subsection{COP of a 2d RT refrigerator}

\begin{figure}[!ht]
\centerline{\psfig{file=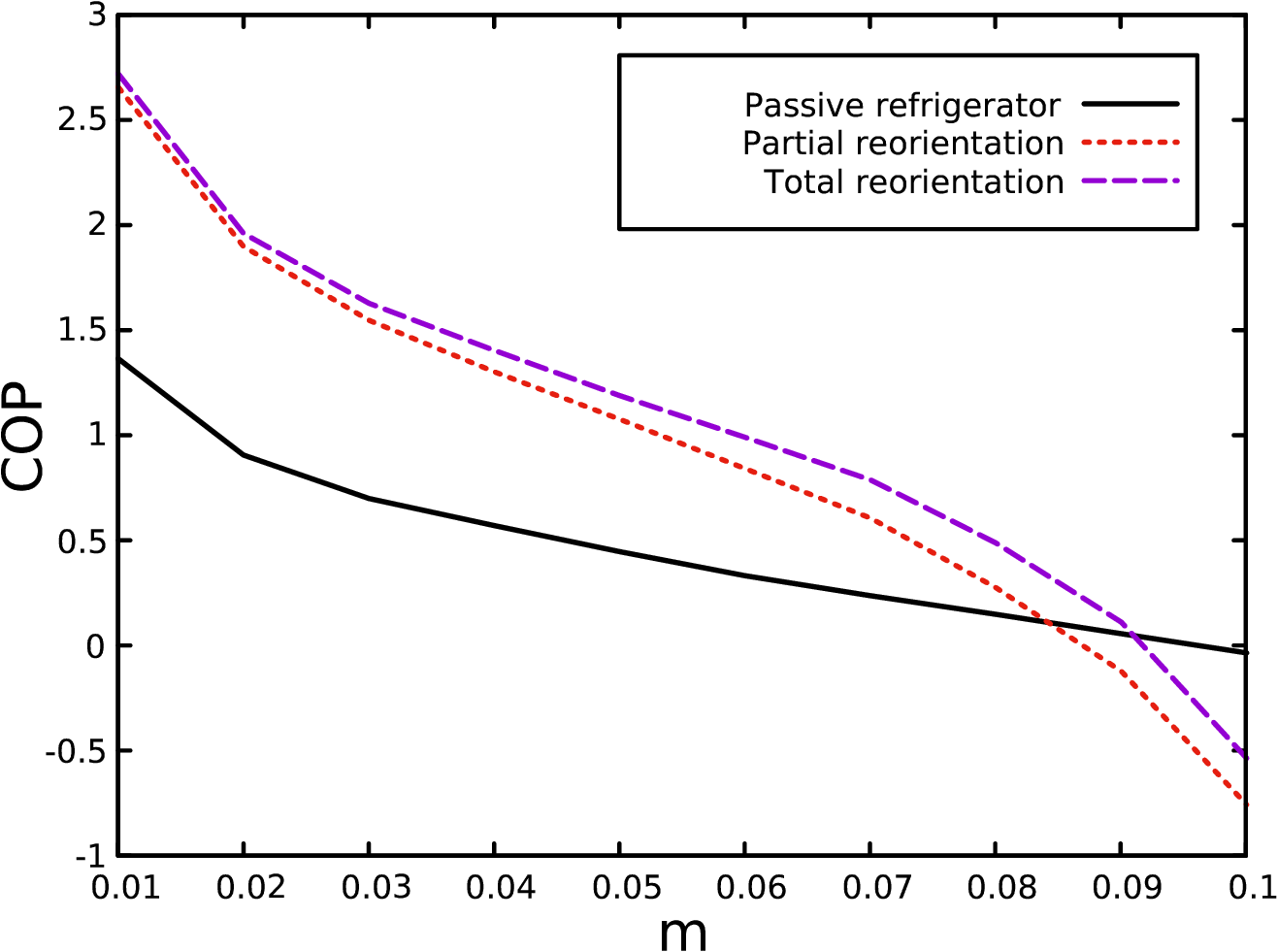,width=2.4in}}
\vspace*{8pt}
\caption{Plots showing the variation in the COP with mass for (a) a passive (black solid line), (b) an RT refrigerator (partial reorientation, red dotted line) and (c) an RT refrigerator (total reorientation, magenta dashed lines) in two dimensions. The parameters used are: $F=0.1$ (for active particle), $D_c=1,D_h=1.25,~k_0=50, \tau_p=0.2$.}
\label{fig:m_cop_comp_2d}
\end{figure}

We now extend our study to that of 2d RT refrigerators (see section \ref{sec:2d_model}). Figure  \ref{fig:m_cop_comp_2d} shows the variations in COP with particle mass for a 2d passive refrigerator as well as for the 2d RT refrigerator (partial and total reorientation). It is apparent that the behaviour resembles that of the 1d refrigerator (see figure \ref{fig:RT1_cop_m}). The total reorientation case seems to offer an advantage over the partial reorientation case, although the differences in COP are not significant. Similar qualitative behaviours are obtained when the activity is present in the full cycle, which we have checked separately. 

\begin{figure}[!ht]
\centering     
\subfigure{\psfig{file=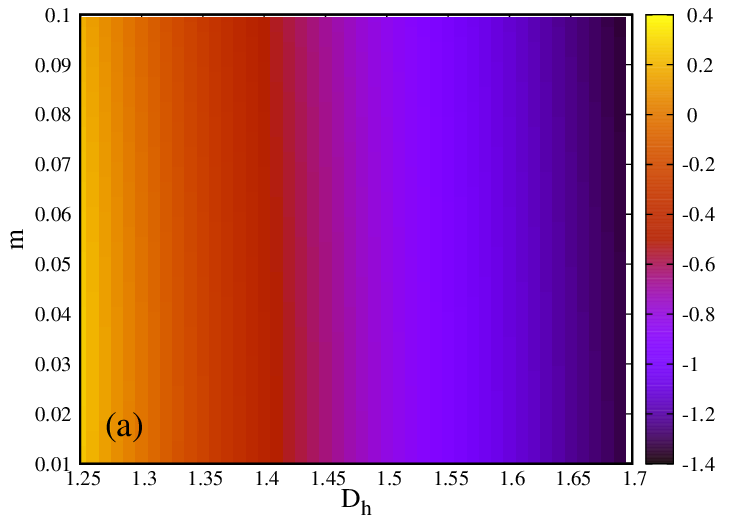,width=2.4in}}
\subfigure{\psfig{file=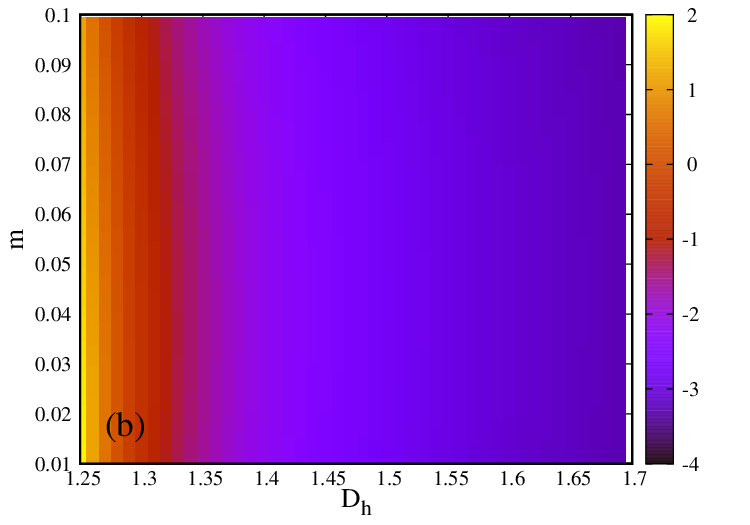,width=2.4in}}
\vspace*{8pt}
\caption{(a)Phase plot showing variation of COP of 2d passive refrigerator with mass and the noise strength of hot bath.(b)Similar phase plot for 2d RT active refrigerator. The parameters are: $D_c=1,~k_0=50,\tau_p=0.2$.}
\label{fig:RT2_m_Dh}

\end{figure}

Let us now compare the functional dependence of the COP on mass and the temperature of the hot bath, for the passive and the active 2d refrigerators (see figure \ref{fig:RT2_m_Dh}). It can be observed that like in the 1d case, we generally have higher values of COP when the refrigerator is active. However, the parameter range within which the RT system acts as a refrigerator ($\av W>0$, $\av{Q_c}<0$) is smaller than in the case of the passive one.

\begin{figure}[!ht]
\centerline{\psfig{file=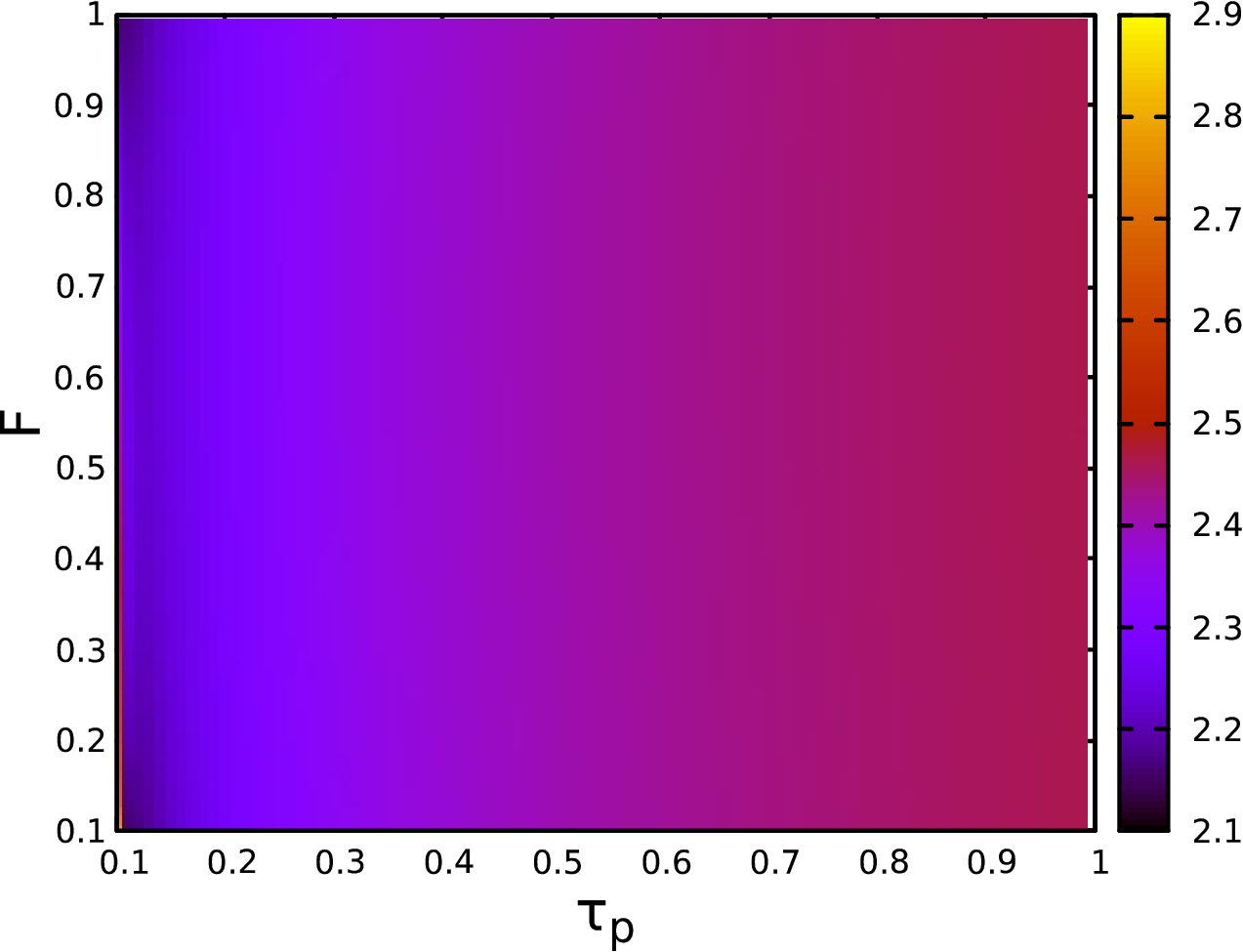,width=2.4in}}
\vspace*{8pt}
\caption{Phase plot showing the dependence of COP of 2d RT refrigerator on the parameters $F$ and $\tau_p$. The parameters are: $D_c=1,~k_0=50,~m=0.01$.}
\label{fig:comp_F_taup_2d}

\end{figure}

We finally discuss about the effect of changes in the active force $F$ and persistence time $\tau_p$ on the performance of the 2d refrigerator. The corresponding phase plot is shown in figure \ref{fig:comp_F_taup_2d}. We find that the non-monotonicity that was present for small values of persistence time for the 1d refrigerator (see figure \ref{fig:comp_F_taup}) does not show up in the 2d case. Instead, to obtain better performance, one should increase the persistence time . 

\section{Conclusions} \label{sec:conclusions}

Microscopic thermal machines have been studied extensively in recent years. Modelling of such machines using a trapped particle in a confining potential has been studied. Recent works show that if a heat engine at such small scales uses an active Ornstein-Uhlenbeck particle as its working system, it can lead to a higher efficiency than that using a passive Brownian particle. We study a microscopic Stirling engine using an active particle, but the activity is introduced by run-and-tumble motion of the particle. In this case,  we find that the efficiency is less than that of a passive engine. 
However, if we apply the refrigerator protocol, then in a suitably chosen parameter regime, the RT refrigerator can give rise to a higher coefficient of performance as compared to its passive counterpart. Both efficiency (for the engine) and COP (for the refrigerator) are higher when the particle mass becomes smaller. An RT motion that involves partial reorientations at the tumble events gives only a minor improvement over that involving total reorientations. For the one-dimensional refrigerator, there is a clear non-monotonic behaviour of the COP when plotted as a function of the active force. It also becomes evident that depending on the nature of the activity (like active Ornstein-Uhlenbeck particle, active Brownian particle, run-and-tumble particle, etc.), the qualitative properties of the nonequilibrium thermal machines are required to be analyzed on a case-by-case basis.

\appendix

\section{Cosine distribution}  

\begin{figure}[!ht]
\centerline{\psfig{file=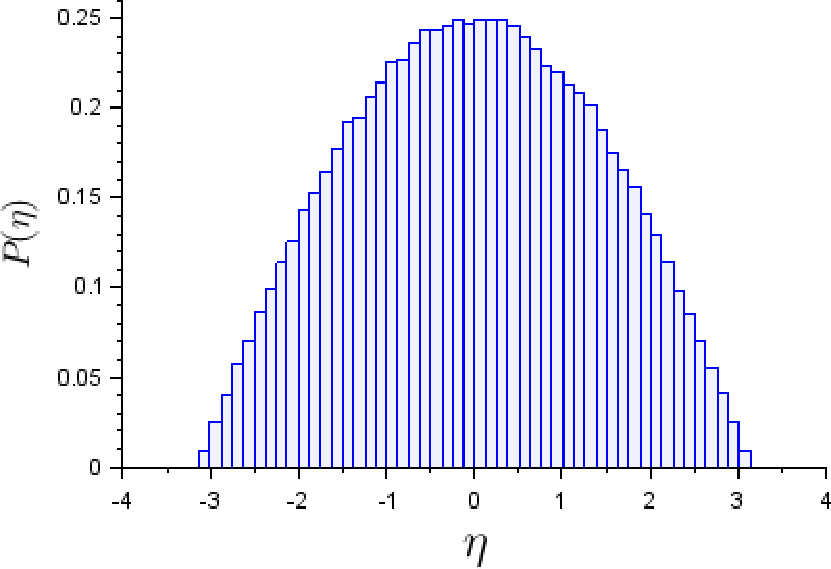,width=3.0in}}
\vspace*{8pt}
\caption{Distribution of tumble angle $\eta$.}
\label{fig:dist}

\end{figure}

To generate the distribution, we have used the technique of inverse transform sampling (ITS) \cite{luc1986}. The cumulative density function (CDF) for $P(\eta)$ is given by
\begin{align}
 F(\eta) &= \frac{1}{4}\int_{-\pi}^\eta \cos\left(\frac{\eta'}{2}\right)~d\eta' = \frac{1}{2}\left[1+\sin\left(\frac{\eta}{2}\right)\right].
\end{align}
Its inverse is given by 
\begin{align}
 \eta&= 2\sin^{-1}[2F(\eta)-1].
\end{align}
Following the method of ITS, we define the random variable in the $i^{\mathrm{th}}$ iteration to be
\begin{align}
 \eta_i &= 2\sin^{-1}(2u_i-1), 
\end{align}
where $u_i$ is a uniform random number in the range [0,1]. The figure \ref{fig:dist} shows that the distribution of $\eta_i$ indeed follows the desired form.

\section{Variance of RT particle in presence of thermal noise}


We follow the analytical technique implemented in \cite{condat2005}
Consider the equation 
\begin{align}
 m\dot v &= -\gamma v + (\sqrt{D})\xi(t).
\end{align}
which is interrupted by tumbles.  In absence of tumble events, the velocity correlation is given by
\begin{align}
\langle v_x(t_1)v_x(t_2)\rangle &= \langle v_y(t_1)v_y(t_2)\rangle = \frac{D}{2m\gamma}e^{-\gamma |t_1-t_2|/m}; 
\end{align}
The variance in absence of tumble events is therefore given by
\begin{align}
 \langle x^2(t)\rangle &= \int_0^t dt_1 \int_0^t dt_2 \left<v_x(t_1)v_x(t_2)\right> = \langle y^2(t)\rangle.
\end{align}
The variance in absence of tumble events is given by
\begin{align}
 \sigma(t) &\equiv \langle x^2(t)\rangle + \langle y^2(t)\rangle = \frac{2D}{\gamma^2}t - \frac{2Dm}{\gamma^3}\left(1-e^{-\gamma t/m}\right).
\end{align}
Next, let us include tumble events. If the probability of a run duration $\tau_r$ is given by $P(\tau_r)$, then the probability that no tumble has taken place in time $t$ is given by
\begin{align}
 Q(t) &= 1-\int_0^t P(t')dt',
\end{align}
so that the modified variance in the presence of tumble events is
\begin{align}
\av{r^2(t)} &= Q(t)\sigma(t) + \int_0^t P(t')[\sigma(t')+\av{r^2(t-t')}]dt'.
\end{align}
The first term on the RHS is the contribution from  runs that are uninterrupted by tumble events, the second term is that of first run and the third is due to subsequent evolution. If we define the function
\begin{align}
 f(t) &= Q(t)\sigma(t) + \int_0^t P(t')\sigma(t')dt',
\end{align}
then performing Laplace transform on both sides, we get
\begin{align}
 \tilde\sigma(s) &= \frac{\tilde f(s)}{1-\tilde P(s)}.
\end{align}
Finally, performing the inverse Laplace transform, we obtain eq. \eqref{var_condat}.

\section{Velocity correlations}

\begin{figure}[!ht]
\centering     
\subfigure{\psfig{file=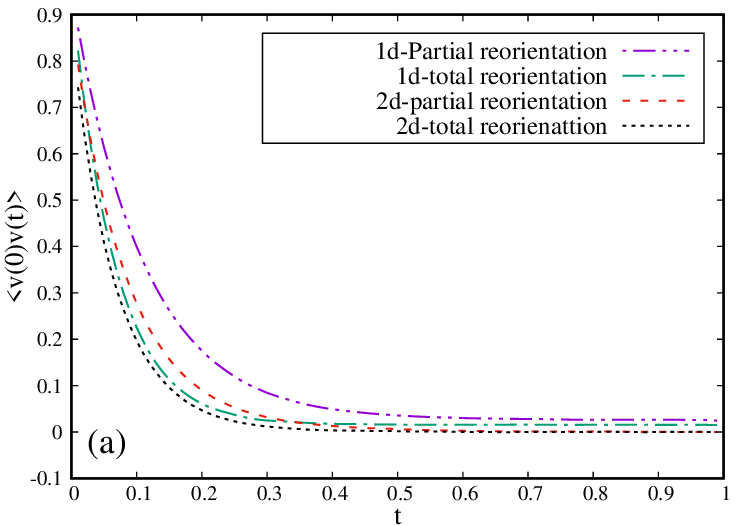,width=2.45in}}
\subfigure{\psfig{file=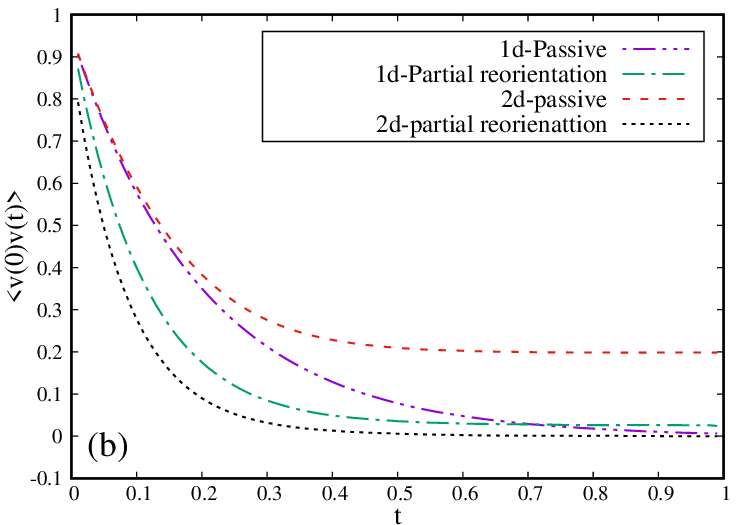,width=2.45in}}
\vspace*{8pt}

\caption{(a)Velocity correlations $\langle v_x(0)v_x(t)\rangle$ as a function of time for a 1d and a 2d RT particle in the case of total and partial reorientations. The parameters used are $m=0.2,~\tau_p=0.1,~F=0.1,~D=0.01,~k_0=0$.(b)Velocity correlations for 1d and 2d passive and active particle with partial reorientation. Parameters used are same as above.}
\label{fig:velcorr}

\end{figure}

Intuitively, we would expect a partial reorientation to correspond to a higher activity as compared to total reorientation, since the former imparts a stronger persistence in the motion of the particle. In order to verify this, we have plotted the velocity correlations along $x$-direction $\langle v_x(0)v_x(t)\rangle$  obtained for a particular set of parameters (as mentioned in the figure caption)  in figure \ref{fig:velcorr} as a function of time $t$. Figure \ref{fig:velcorr}(a) compares the velocity correlations for 1d and 2d RT particles, both for total and partial reorientations. As expected, the correlations are higher for partial reorientation in each case. Figure \ref{fig:velcorr}(b) compares the velocity correlations of the 1d and 2d particles undergoing partial reorientations, with those of the corresponding passive particles. The correlations are found to be generally weaker for the RT particles, due the degradation in the correlations caused by tumble events.

\section{Brief description of engine using an Active Ornstein-Uhlenbeck particle}

We have compared our results with those of \cite{lahiri2020}, where the overdamped AOUP has been used. Such a particle follows a Langevin equation where, in addition to the thermal noise, there is an exponentially correlated noise that leads to persistence in its motion. The equations of motion for the AOUP are given below for the expansion and compression steps respectively:
\begin{align}
 \gamma \dot x &= -k_{\rm exp}(t)x + (\sqrt{D_h})\xi(t), \hspace{1cm} 0\le t<\tau/2;\nonumber\\
 \gamma \dot x &= -k_{\rm com}(t)x + (\sqrt{D_c})\xi(t) + (\sqrt{D_\eta/\tau_\eta})\eta(t), \nonumber\\
 &\hspace{5cm} \tau/2\le t <\tau; \nonumber\\
 \tau_\eta\dot\eta &= -\eta + (\sqrt{2\tau_\eta})\xi_\eta(t).
\end{align}
Here, $\tau_\eta$ is the decay time of the exponentially correlated noise $\eta$:
\begin{equation}
 \av{\eta(t)\eta(t')} = e^{-|t-t'|/\tau_\eta}.
\end{equation}
The constant $D_\eta$ determines the strength of the noise (for a fixed value of $\tau_\eta$), which to call the active noise strength. The functions $k_{\rm exp}(t)$ and $k_{\rm com}(t)$, as well as the constants $D_h$ and $D_c$, are the same as given in section \ref{sec:1d model}.


\section*{Acknowledgments}

The authors thank DST-SERB  (grant number \\ECR/2017/002607) for funding. SL thanks S. Chaki for suggesting useful references.





\begin{thebibliography}{99} 
\bibitem{vyas2014}
Y Saadeh and D Vyas,
\newblock {\em Am. J. Robotic Surgery}, {\bf 1}, 4 (2014).

\bibitem{sei08_epl}
T Schmeidl and U Seifert,
\newblock {\em Europhys. Lett.}, {\bf 81}, 20003 (2008).

\bibitem{saha2019}
A Saha and R Marathe,
\newblock {\em J. Stat. Mech: Theor. Exp}, {\bf 2019}, 094012 (2019).

\bibitem{lahiri2020}
A Kumari, P S Pal, A Saha, and S Lahiri,
\newblock {\em Phys. Rev. E}, {\bf 101}, 032109 (2020).

\bibitem{roldan2016}
I A Martinez, E Roldan, L Dinis, D Petrov, J M R Parrondo, and R A Rica,
\newblock {\em Nat. Phys.}, {\bf 12}, 67 (2016).

\bibitem{bechinger2012}
V Blickle and C Bechinger,
\newblock {\em Nat. Phys.}, {\bf 8}, 143 (2012).

\bibitem{lutz2016}
J Ro{\ss}nagel, S T Dawkins, K N Tolazzi, O Abah, E Lutz,
  F S Kaler, and K Singer, 
\newblock {\em Science}, {\bf 352}, 325 (2016).

\bibitem{sood2016} S Krishnamurthy, S Ghosh, D Chatterji, R Ganapathy, and A K Sood, 
\newblock {\em Nat. Phys.}, {\bf 12}, 1134 (2016).

\bibitem{roldan2017}
I ~A Martinez, E Roldan, L Dinis, and R A Rica,
\newblock {\em Soft Matter}, {\bf 13}, 22 (2017).

\bibitem{bender2000}
C M Bender, D C Brody, and B K Meister,
\newblock {\em J. Phys. A}, {\bf 33}, 4427 (2000).

\bibitem{quan2007}
H T Quan, Y X Liu, C P Sun, and F Nori,
\newblock {\em Phys. Rev. E}, {\bf 76}, 031105 (2007).

\bibitem{rana2016}
S Rana, P S Pal, A Saha, and A M Jayannavar,
\newblock {\em Physica A}, {\bf 444}, 783 (2016).

\bibitem{pal2016}
P S Pal, A Saha, and A M Jayannavar.
\newblock {\em Int. J. Mod. Phys. B}, {\bf 30}, 1650219 (2016).

\bibitem{ai2006}
B Q Ai, L Wang, and L G Liu,
\newblock {\em Phys. Lett. A}, {\bf 352}, 286 (2006).

\bibitem{lin2009}
B Lin and J Chen,
\newblock {\em J. Phys. A: Math. Theor.}, {\bf 42}, 075006 (2009).

\bibitem{chen2010}
L Chen, Z Ding, and F Sun,
\newblock {\em Appl. Math. Model.}, {\bf 35}, 2945 (2010).

\bibitem{leo2010}
R ~Di Leonardo, L ~Angelani, D ~Dell’Arciprete, G ~Ruocco,  V ~Iebba, S ~Schippa, M ~P ~Conte, F ~ Mecarini, F ~De   Angelis, and E ~Di Fabrizio,
\newblock {\em PNAS}, {\bf 107}, 9541 (2010).

\bibitem{liang2000}
S Liang, D Medich, D ~M Czajkowsky, S Sheng, J Y
  Yuan, and Z Shao,
\newblock {\em Ultramicroscopy}, {\bf 84}, 119 (2000).

\bibitem{kim2004}
K H Kim and H Qian,
\newblock {\em Phys. Rev. Lett.}, {\bf 93}, 120602 (2004).

\bibitem{kim2007}
K H Kim and H Qian,
\newblock {\em Phys. Rev. E}, {\bf 75}, 022102 (2007).

\bibitem{arxiv2008}
H J Briegel and S Popescu,
 \newblock {\em arxiv/quant-ph:0806.4552}, (2008).

\bibitem{linden2010}
N Linden, S Popescu, and P Skrzypczyk,
\newblock {\em Phys. Rev. Lett.}, {\bf 105}, 130401 (2010).

\bibitem{brunner2014}
N Brunner, M Huber, N Linden, S Popescu, R Silva, , and
  P Skrzypczyk,
\newblock {\em Phys. Rev. Lett.}, {\bf 89}, 032115 (2014).

\bibitem{berg1972}
H C Berg and D A Brown,
\newblock {\em Nature}, {\bf 239}, 500 (1972).

\bibitem{condat2005}
C A Condat, J J\"ackle, and S A Mench\'on,
\newblock {\em Phys. Rev. E}, {\bf 72}, 021909 (2005).

\bibitem{buceta2017}
G Fier, D Hansmann, and R C Buceta,
\newblock {\em Soft Matter}, {\bf 13}, 3385 (2017).

\bibitem{buceta2018}
G Fier, D Hansmann, and R C Buceta,
\newblock {\em arxiv/cond-mat:1802.00269}, (2018).

\bibitem{schnitzer1993}
M J Schnitzer,
\newblock {\em Phys. Rev. E}, {\bf 48}, 2553 (1993).

\bibitem{dhar2019}
A Dhar, A Kundu, S N Majumdar, S Sabhapandit, and
  G Schehr,
\newblock {\em Phys. Rev. E}, {\bf 99}, 032132 (2019).

\bibitem{chaki2019}
S Chaki and R Chakrabarti,
\newblock {\em J. Chem. Phys.}, {\bf 150}, 094902 (2019).

\bibitem{cates2013}
M E Cates and J Tailleur,
\newblock {\em Europhys. Lett.}, {\bf 101}, 20010 (2013).

\bibitem{elgeti2015}
J Elgeti, R G Winkler, and G Gompper,
\newblock {\em Rep. Prog. Phys.}, {\bf 78}, 056601 (2015).

\bibitem{sano2017}
T Manoa, J B Delfauc, J Iwasawaa, and M Sano,
\newblock {\em PNAS}, {\bf 114}, E2580 (2017).

\bibitem{soto2020}
A Villa-Torrealba, C Ch\'avez-Raby, Pablo de~Castro, and
  R Soto,
\newblock {\em Phys. Rev. E}, {\bf 101}, 062607 (2020).

\bibitem{wijland2017_entropy}
R ~Zakine, A ~Solon, T Gingrich, and Fr\'ed\'eric van Wijland,
\newblock {\em Entropy}, {\bf 19}, 193 (2017).

\bibitem{buttinoni2012}
 I Buttinoni, G Volpe, F Kümmel, G Volpe, and C Bechinger,
\newblock {\em J. Phys.: Condens. Matter}, {\bf 24}, 284129 (2012).

 \bibitem{uspal2019}
 W E Uspal,
\newblock {\em J. Chem. Phys.}, {\bf 150}, 114903 (2019).   

\bibitem{berg1974}
D A Brown and H C Berg,
\newblock {\em Proc. Nat. Acad. Sci. USA}, {\bf 71}, 1388 (1974).

\bibitem{man2000}
R Mannela,
\newblock {\em Stochastic process in physics, chemistry and biology},
\newblock edited by J A Freun and T P\"oschel, (Springer-Verlag, Berlin, 2000).

\bibitem{sek98}
K Sekimoto,
\newblock {\em Prog. Theor. Phys. Supp.}, {\bf 130}, 17 (1998).

\bibitem{sekimoto}
K Sekimoto,
\newblock in {\em Stochastic Energetics},
\newblock (Springer-Verlag, Berlin, 2010).

\bibitem{luc1986}
L Devroye,
\newblock in {\em Non-Uniform Random Variate Generation},
\newblock (Springer, New York, 1986).

\end{thebibliography}
\end{document}